\documentclass{pasa}%
\usepackage[T1]{fontenc}
\usepackage{graphicx}

\title[Incompressible models for pulsars] 
{Incompressible analytical models for spinning-down pulsars} 

\author[Giliberti et al.]{
E. Giliberti$^{1,4}$\thanks{elia.giliberti@unimi.it},
M. Antonelli$^{2}$\thanks{mantonelli@camk.edu.pl},
G. Cambiotti$^{3}$ 
and P.M. Pizzochero$^{1,4}$ 
\affil{$^{1}$Dipartimento di Fisica, Universit\`a degli Studi di Milano, Via Celoria 16, 20133,Milano, Italy}
\affil{$^{2}$Nicolaus Copernicus Astronomical Center, ul. Bartycka 18, 00-716 Warsaw, Poland\\
$^{3}$Dipartimento di Scienze della Terra, Universit\`a degli Studi di Milano, Via Cicognara 7, Milano, 20129, Italy}
\affil{$^{4}$Istituto Nazionale di Fisica Nucleare, sezione di Milano, Via Celoria 16, 20133 Milano, Italy}
}

\usepackage{aas_macros}
\usepackage{hyperref} 
\hypersetup{colorlinks,citecolor=blue,linkcolor=blue,urlcolor=blue}

\begin{document}

\begin{frontmatter}
\maketitle

\begin{abstract}
We study a class of Newtonian models for the deformations of non-magnetized neutron stars during their spin-down. The models have all an analytical solution, and thus allow to understand easily the dependence of the strain on the star's main physical quantities, such as radius, mass and crust thickness. In the first ``historical'' model the star is assumed to be comprised of a fluid core and an elastic crust with the same density. We compare the response of stars with different masses and equations of state to a decreasing centrifugal force, finding smaller deformations for heavier stars: the strain angle is peaked at the equator and turns out to be a decreasing function of the mass.We introduce a second, more refined, model in which the core and the crust have different densities and the gravitational potential of the deformed body is self-consistently accounted for. Also in this case the strain angle is a decreasing function of the stellar mass, but its maximum value is at the poles and is always larger than the corresponding one in the one-density model by a factor of two. 
Finally, within the present analytic approach, it is possible to estimate easily the impact of the Cowling approximation: neglecting the perturbations of the gravitational potential, the strain angle is 40\% of the one obtained with the complete model.
\end{abstract}


\end{frontmatter}


\section{INTRODUCTION}

The long-term evolution of a neutron star (NS) can be driven by spin-down, accretion of matter or external
forces, like the tidal force due to the presence of a close companion or electromagnetic strains arising from strong internal magnetic fields. 
According to the current understanding of matter properties at supra-nuclear densities, neutron star structure involves a superfluid core surrounded by a floating hard crust \citep[see e.g.][for a comprehensive review]{chamel_livingreview}. All the aforementioned processes may lead to deformation of the crust, measured by considering the displacement with respect to an initial equilibrium state \citep[usually assumed to be a stationary fluid configuration,][]{love59}.

Stellar crust breaking is thought to be a key aspect for understanding several phenomena linked to the astrophysical phenomenology of neutron stars. In fact, starquakes are promising candidates as trigger mechanism for both  glitches in radio pulsars \citep{baym1969,rudermanII1991} and flares in magnetars \citep{thompson1995,cheng1996}. The crust breaking hypothesis is studied also for its possible role on NSs precession \citep{link1998,cutler2003} and on gravitational waves emission \citep{usho2000,haskell2006}. For all of these phenomena the crust acts as an elastic layer that can store stress during time \citep{baym1971}, until it reaches a critical threshold defined by the breaking strain value of the material (see e.g. \cite{failure_book} for a extensive discussion on failure criteria). 

Since the seminal work of Love on the theory of the elastic response of a homogeneous, self-gravitating body of nearly spherical shape \citep{love59}, only two analytic models describing deformations of a neutron star have been  presented to date: in order to investigate the qualitative features of the growth of strain in a neutron star as it spins down under its external torque, \citet{baym1971} modeled the star as a self-gravitating, incompressible, homogeneous elastic sphere of constant shear modulus. This model allowed \citet{link1998} to conclude that the failure of the crust is likely to occur near the equator, with interesting consequences on the magnetic field and the evolution of the electromagnetic braking torque of pulsars.
Later, \citet{franco2000} refined the model of Baym and collaborators by considering a star composed of a fluid (i.e. a substance of null shear modulus) core and a solid crust with the same constant densities, finding that the introduction of the fluid core does not affect the conclusion that the crust is likely to break near the equator. We will refer to this model with discontinuous shear modulus at the core-crust interface as FLE model.

In this paper we study the behavior of FLE-like models in the limits of astrophysical interest. Parameters, like the radius, crust thickness and mass, are fixed by considering realistic equations of state of dense matter in neutron stars. After a revision of the original FLE approach, we introduce in Section 4 a more refined analytic model that is based on the set of ideas presented in \citet{sabadini82} and \citet{sabadini1996} for the study of the a viscoelastic Earth, here adapted to the neutron star scenario in the elastic limit. 
Our new model incorporates the gravitational potential in a fully consistent way and allows to solve exactly the problem of an auto-gravitating rotating body with a fluid interior and an elastic envelope, with two different densities. 
A comparison is made between the different models by focusing on the spin-down of a pulsar between two glitches, in terms of the calculated strains in the crust. 

The exact formulae for the displacement field given by these analytical approaches are given in the Appendix. The important effects due to the finite compressibility and stratification of matter, not included in these models, are considered in the more general approach described in \citet{Giliberti2018}. However, compressible and self-gravitating models are considerably more complex and need a certain amount of numerical work to be solved, so that the kind of two-density models studied here represent an improvement of the approach pioneered by \citet{baym1971}, without losing the possibility to obtain closed solutions for the displacement field ready to be used for astrophysical estimates.    


\section{ELASTICITY}

\begin{figure}
\centering
\includegraphics[width=0.7\columnwidth]{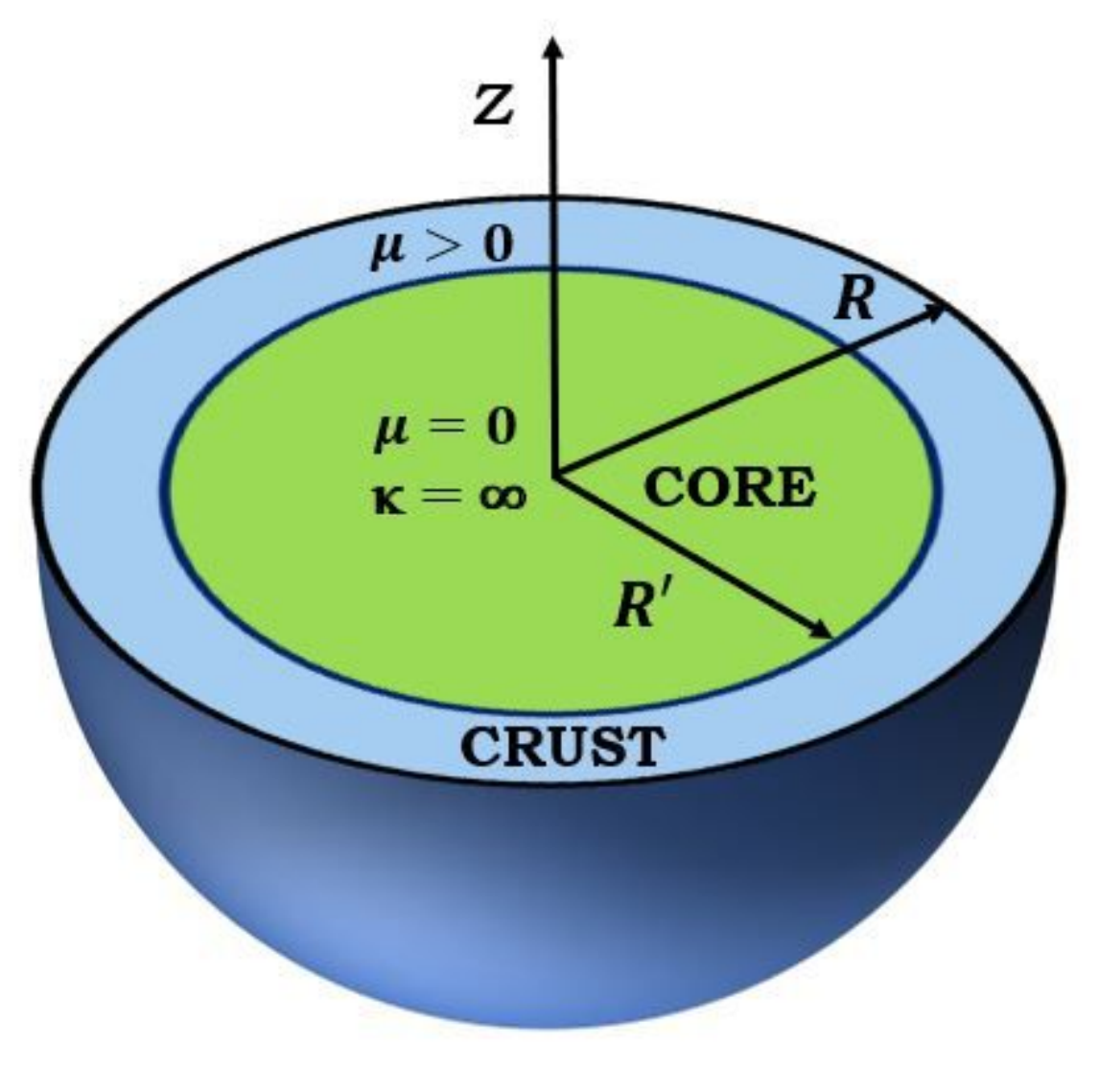}
\caption{
Sketch of the idealized non-rotating neutron star structure used in the present study (not in scale). The spherical configuration of radius $R$ consists of two regions separated at $r=R'$: the core, where the shear modulus $\mu$ is null, and the crust, where $\mu>0$. Since all the models discussed in the present work are incompressible, the bulk  modulus $\kappa$ is infinite everywhere.
}
\label{fig:FIGURA STELLA}
\end{figure}

Let us consider a non-rotating, unstrained NS of radius $R$ and core radius $R'$. This will be our initial configuration, sketched in Fig \ref{fig:FIGURA STELLA}. For a given equation of state (EoS), the values of $R$ and $R'$ can be found by solving the problem of hydrostatic equilibrium. 

We define $\boldsymbol{x}$ to be the position  of a portion of matter in the initial configuration, $\boldsymbol{r}(\boldsymbol{x})$ the new position of the same portion of matter in the configuration of the rotating body and the local displacement $\boldsymbol{u}$ as the difference of the previous two, i.e.
\begin{equation*}
\boldsymbol{r}(\boldsymbol{x}) \,  = \, \boldsymbol{x} + \boldsymbol{u}(\boldsymbol{x}) \, .
\end{equation*}
Therefore, the deformation of the star due to rotation with respect to the initial configuration can be described by the strain tensor $u_{ij}$ \citep{love59,landau_book}.
%
%

Since the centrifugal force has azimuthal symmetry, the displacement cannot depend on the azimuthal angle, so that $\boldsymbol{u}=\boldsymbol{u}(r,\theta)$, where $r$ is the usual radial distance from the center of the star and $\theta$ is the colatitude angle. 
For the determination of the crust breaking we will use the Tresca criterion: 
introducing the \emph{strain angle} $\alpha$ as the difference between the local maximum, $\epsilon_{max}$, and minimum,  $\epsilon_{min}$, eigenvalues of the strain tensor,
\begin{equation}
\alpha(r,\theta)=\epsilon_{max}(r,\theta)-\epsilon_{min}(r,\theta) \, ,
\label{definizione alpha}
\end{equation}
the material response to strains ceases to be elastic first in the regions of the crust where the strain angle is maximum. The material locally cracks around the zone where the strain angle is peaked when
\begin{equation*}
\alpha_{max}=\frac{\sigma_{max}}{2} \, ,
\end{equation*}
where $\alpha_{max}$ is the maximum of the function $\alpha(r,\theta)$ in the range $R'\leq r \leq R$ and $0 \leq\theta\leq\pi$, and $\sigma_{max}$ is a property of the material called \emph{breaking strain} (see e.g. \cite{failure_book}).

To date the value of the breaking strain in a neutron star crust is poorly known. Estimates of $\sigma_{max}$  span from values of the order of $10^{-2}\div10^{-1}$ \citep{horo2009,baiko2018} found by means of microscopic calculations to the much lower values of $10^{-5}\div 10^{-3}$ used in macroscopic\footnote{ 
It is currently debated whether or not microscopic scale failure (usually investigated by means of molecular dynamics to signify failure when bonds are distorted beyond their limits) is predictive of macroscopic failure, as discussed in \citet{failure_book}.} 
models of glitches and flares \citep{rudermanII1991,thompson1995}. 
Because of these theoretical uncertainties, we will assume that a realistic value $\sigma_{max}$ for a macroscopic portion of crustal matter is the range $10^{-5}\div 10^{-1}$. 

Since crustal matter is likely to be an isotropic bcc polycrystal, a single effective shear modulus $\mu$ and the bulk modulus $\kappa$ are expected to be sufficient to express stresses in terms of strains via Hooke's law \citep[see ][and references therein]{chamel_livingreview}.

The shear modulus, in particular, can be calculated as a function of the crustal composition \citep{Ogata1990,horo2008}. More recently, \citet{caplan18} performed classical molecular dynamics simulations where  a sample of nuclear pasta at the bottom of the inner crust is deformed; they simulate idealized samples of nuclear pasta and describe their breaking mechanism, finding that nuclear pasta may be the strongest layer of a neutron star, with a shear modulus of $\mu \sim 10^{30}\,$erg/s.

If matter is at equilibrium, the crustal composition can be thought as a function of the total density $\rho$, and thus $\mu=\mu(\rho)$. However, continuous stratification introduces a considerable complication that can be dealt with the more refined approach proposed in \citet{Giliberti2018}, which has to be solved numerically (see also \cite{usho2000}). 
Therefore, in order to obtain exactly solvable analytic models, in the following we will study only idealized stellar structures with homogeneous layers, implying also constant shear and bulk modulus, as sketched in Fig \ref{fig:FIGURA STELLA}. 
In the present work we consider that the Cauchy stress tensor is given by
\[
T_{ij} 
\, = \, 
-p \, \delta_{ij} \, + \, \sigma_{ij} 
\, = \,
-p \, \delta_{ij} \, + \, \mu \, u_{ij}
\, ,
\]
where $p$ is the local pressure at equilibrium and $\sigma_{ij}$ is the material incremental stress, where we made use of the incompressibility assumption $u_{ii}=0$. The shear modulus $\mu$ is constant in the crust and it is set to zero in the core.


\section{FLE MODEL}

In this section we study in detail the FLE model \citep{franco2000}, where the star is described as a homogeneous body, with a fluid core and an elastic crust with the same density.
We are interested in calculating the displacement field $\boldsymbol{u}$ between a configuration rotating with velocity $\Omega$ and the one rotating at $\Omega-\delta \Omega$, where $\delta\Omega\,>0$ for a spinning down pulsar.
The non-rotating configuration is known for our elastic star since it coincides with the one given by the usual hydrostatic equilibrium for a fluid. Thanks to the assumed linearity of the problem, we calculate the displacements $\boldsymbol{u}_\Omega $ due to the spin-up of a spherical configuration to a rotating one having centrifugal potential proportional to $\Omega^2$; then, the desired displacement between the two rotating configurations is given by
\[
\boldsymbol{u} 
\, = \,   
\boldsymbol{u}_\Omega  -  \boldsymbol{u}_{\Omega-\delta \Omega}
\,   
\propto \, 
\delta \Omega \, \Omega \, . 
\] 
Clearly this procedure gives results identical to the method described by \cite{franco2000}, according to which the displacement field $\boldsymbol{u}$ up to the linear order in $\delta\Omega$ is  
\begin{align}
\begin{split}
u_{r} (r,\theta)
& =
\left(ar-\frac{1}{7}Ar^{3}-\frac{1}{2}\frac{B}{r^{2}}+\frac{b}{r^{4}}\right) P_{2}
\\
u_{\theta}(r,\theta) 
& =
\left(\frac{1}{2}ar-\frac{5}{42}Ar^{3}-\frac{1}{3}\frac{b}{r^{4}}\right)\frac{dP_{2}}{d\theta}
,
\label{SPOSTAMENTO FLE}
\end{split}
\end{align}
where $P_{2}=\frac{1}{2}\left(3 \cos^{2}\theta-1\right)$ is the second Legendre polynomial of argument $\cos \theta$. The four coefficients $a$, $b$, $A$ and $B$ are fixed by four boundary conditions, two at the core-crust transition radius $r=R'$ and two at the star's surface $r=R$. At both these interfaces we have to require the continuity of radial stresses, $T_{rr}=-p+\mu u_{rr}$, and that $\sigma_{r\theta}=0$, since both the fluid core and the vacuum outside the star cannot support shears.
It is useful to introduce the sound speed in the crust of transverse waves, $c_{t}=\sqrt{\mu/\rho}$, and the usual Keplerian velocity, $v_{K}=\sqrt{GM/R}$, so that the four boundary conditions read
\begin{align}
\begin{split}
& a-\frac{8}{21} A R^{2} - \frac{B}{2R^{3}}+\frac{8}{3}\frac{b}{R^5}
= 0
\\ 
& a-\frac{8}{21}A R'^{2} - \frac{B}{2R'^{3}}+\frac{8}{3}\frac{b}{R'^5}
= 0
\\
& f'(R)+\frac{1}{5}\frac{v_{K}^{2}}{c_{t}^{2}}\frac{f(R)}{R}-\frac{1}{3}\frac{\Omega\delta\Omega}{c_{t}^{2}}R^{2}
= -\frac{A R^{2}}{2}
\\
& f'(R') 
= -\frac{1}{2}\left(A R'^2 + \frac{B}{R'^{3}}\right) \, . 
\label{CONDIZIONI DI BORDO}
\end{split}
\end{align}
Using the definition \eqref{SPOSTAMENTO FLE} and the boundary conditions \eqref{CONDIZIONI DI BORDO}, the four coefficient $a, b,A,B$ are obtained with straightforward algebra. 
It seems more useful to rewrite the displacement \eqref{SPOSTAMENTO FLE} as 
\begin{align}
\begin{split}
u_{r}
& =
\frac{\left(\Omega\delta\Omega\right)\,R^{3}}{Q(c_{t},v_{K},L)}
\left(\frac{\tilde{a} r}{R}-\frac{\tilde{A} r^{3}}{7 R^{3}}-\frac{\tilde{B} R^{2}}{2 r^{2}}+\frac{\tilde{b} R^{4}}{r^{4}}\right)
P_{2}
\\
& =f(r)P_{2}
\\
u_{\theta}
& =
\frac{\left(\Omega\delta\Omega\right)\, R^{3}}{Q(c_{t},v_{K},L) }
\left(\frac{\tilde{a} r}{2 R}-\frac{5 \tilde{A} r^{3}}{42 R^{3}}-\frac{\tilde{b} R^{4}}{3 r^{4}}\right)
\frac{dP_{2}}{d\theta}
\\
& = g(r)\frac{dP_{2}}{d\theta} ,
\label{Spostamento Adimensionale}
\end{split} 
\end{align}
where the tilde superscript indicates that now the coefficients are dimensionless: all the dependence on physical parameters has been included into the pre-factors ($Q$ is a squared velocity built with the typical scales of the problem). Therefore, $\tilde{a}$, $\tilde{A}$,  $\tilde{b}$ and  $\tilde{B}$  are functions of the parameter $ L = R' / R $ only; clearly, the limits $L=0$ and $L=1$ describe a completely solid star and a completely fluid star respectively. The explicit form of the parameters present in \eqref{Spostamento Adimensionale}, including $Q$ , is given in Appendix B.

Finally we introduce the parameter $\chi$, defined as the ratio
\begin{equation}
    \chi\, = \, \frac{c_t}{v_K} \, \ll \,  1 
    \, ,
    \label{DEFINIZIONE CHI}
\end{equation}
which, according to current estimates of $\mu \sim 10^{28}$ erg/cm$^3$, is expected to be much less than unity in the whole crust \citep[see e.g. Fig 7 of ][]{zdunik08}. Interestingly, as noted by several authors \citep[see e.g.][]{haskell2006,bastrukov07}, the speed of transverse elastic shear waves $c_t \sim 10^{8}\,$cm/s is rather constant (within a factor of $2$) throughout the crust, so that we expect $\chi \sim 10^{-2}$.

%

\subsection{Parametric study of the FLE model}

In their original work, \citet{franco2000} considered a ``standard'' NS of mass $M=1.4 M_{\odot}$, $R=10\,$km and $R'=0.95\,R$ (i.e. $L=0.95$ according to the present notation), as benchmark stellar configuration. Here, we extend their analysis investigating the behavior of the FLE model as a function of
the star's parameters: radius, mass and crust thickness.

Let's focus for a moment on the displacement \eqref{Spostamento Adimensionale}. We can define a dimensionless weight factor $W$ 
\begin{equation}
    W=\frac{\left(\Omega \, \delta\Omega\right)\,R^{2}}{Q(c_t,v_K,L)}
\end{equation}
Using the smallness of the $\chi$ parameter we can write, see also equation \eqref{Q FLE} in the Appendix,
\begin{equation}
    Q\propto v_K^{2}.
\end{equation}
%
%
%
Fixing $\Omega \, \delta\Omega$ to some constant in order to remove the dependence on the rotational parameters of a particular pulsar, we see that for the FLE model
%
%
%
\begin{equation}
   W 
    \propto 
    \frac{R^{2}}{v_K^{2}}=
    \frac{R^3}{GM}
    \propto
    \frac{1}{\rho} \,  .
    \label{W caso generale}
\end{equation}
Therefore, the denser the star, the smaller the displacement, as can be easily seen from equation \eqref{Spostamento Adimensionale}.
%

\begin{figure}
		\includegraphics[width=\columnwidth]{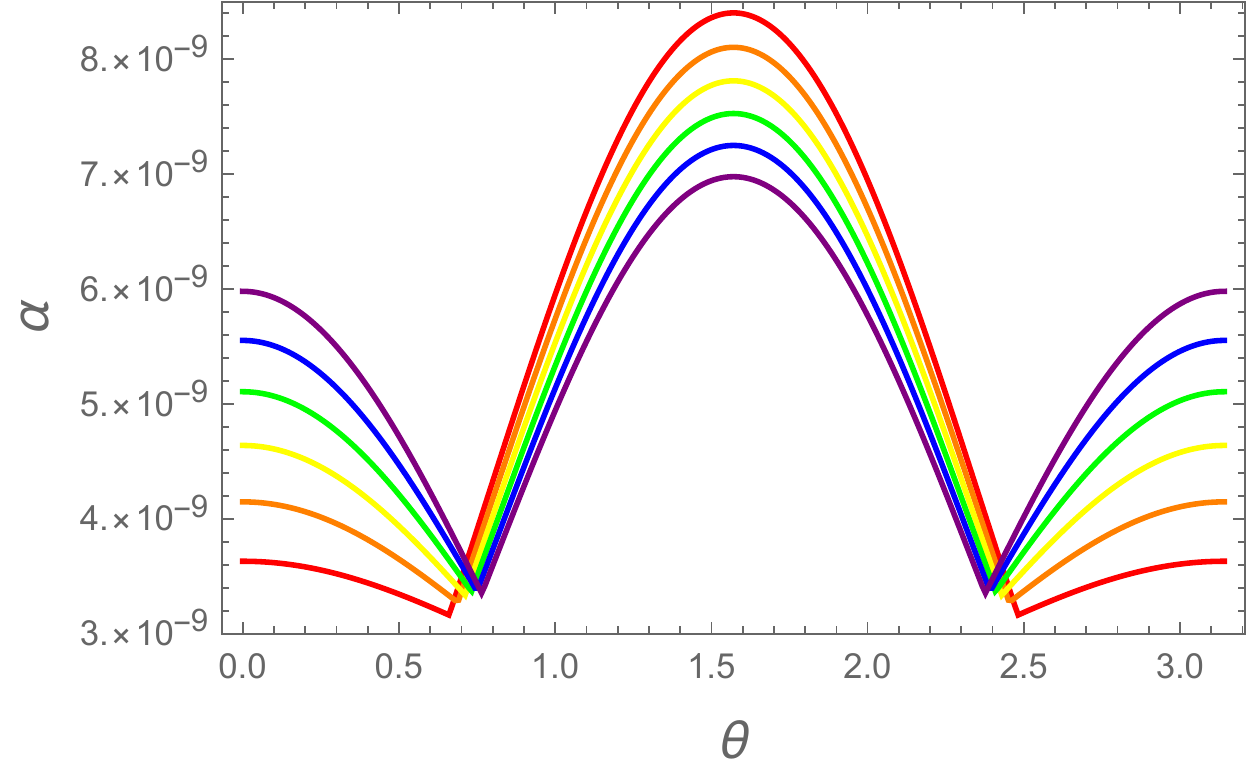}
    \caption{
    Strain angle as a function of the colatitude $\theta$ for the FLE model 
    and fixed benchmark values $M=1.4M_{\odot}$, $R=10\,$km, $L=0.95$. 
    The strain angle is calculated for different values of the radius: $r=R$ (purple), $r=0.99 R$ (blue), $r=0.98 R$ (green), $r=0.97 R$ (yellow), $r=0.96 R$ (orange) and $ r=0.95R$ (red). 
    We used $\Omega\delta\Omega =1\,$rad$^2$/s$^2$.
    }
    \label{fig:STRAIN TUTTO VARIABILE DIVERSI RAGGI}
\end{figure}

In order to better understand this behavior, the strain angle value $\alpha$ may be calculated changing the parameters one by one, keeping fixed all the others.  For comparison purposes, we set the relative extension of the core to be $L=0.95$, the same used in the original FLE study. 

Since the displacements, and thus the strain, are proportional to the actual spin down that occurred between the two configurations, we set the pre-factor $\Omega\delta\Omega$ equal to one\footnote{
Incidentally, our choice to set $\Omega\delta\Omega= 1\,$rad$^2$/s$^2$ in all the calculations of the plotted strains is not so distant from the fiducial value $\Omega\delta\Omega \approx 0.6\,$rad$^2$/s$^2$ that we will use for the Vela pulsar (B0833-45) in the next section (see table \ref{tab:table 1}).
} 
for simplicity. Therefore, to calculate the deformation for a certain star it is just sufficient to multiply the desired quantities for the actual parameter $\Omega \, \delta\Omega$.

Let us first observe that, for given mass, radius and crust thickness of the star, the strain angle $\alpha$ is a decreasing function of $r$, as can be seen in Fig. \ref{fig:STRAIN TUTTO VARIABILE DIVERSI RAGGI}. Hence, we expect that, if the crust breaks, the failure threshold  will be reached first at the crust-core interface, near the equatorial plane (i.e. $\theta \approx \pi/2$ in our coordinates). 

\begin{figure}
		\includegraphics[width=\columnwidth]{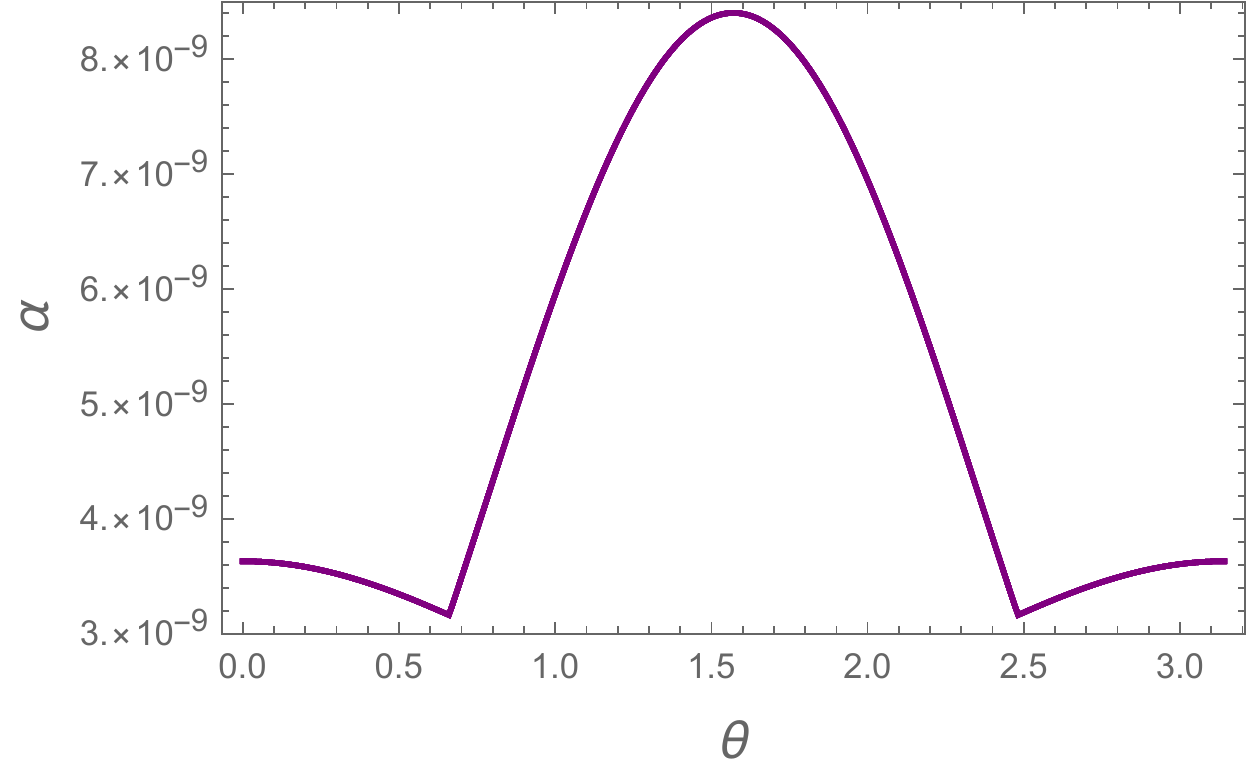}
    \caption{
    Strain angle $\alpha$ of the FLE model restricted to the spherical shell $r=R'$ as a function of the colatitude $\theta$. The crust thickness parameter is fixed to $L=0.95$ but we consider two extreme values of the stellar radius: $R=10\,$km and  $R=20\,$km. The corresponding two masses $M$ are fixed by the constraint that the average density of both configurations is $\rho=6.6\times10^{14}\,$g/cm$^{3}$. The two curves appear to be superimposed in the graph.
    }
    \label{fig:STRAIN TUTTO VARIABILE DENSITA' FISSATA}
\end{figure}

Moreover, Eq. \eqref{W caso generale} suggests us that it can be interesting to compare the strains of stars all having the same average density $ \rho \, = \, 3 M/(4\pi R^{3}) $ but different radii and masses.
We find that, as long as the density is taken constant but the mass and the radius vary, the strain is nearly unchanged. An example of this is shown in Fig \ref{fig:STRAIN TUTTO VARIABILE DENSITA' FISSATA}, where we fixed $\rho$ to be the average density of a star of $1.4 \, M_\odot$ and $R=10\,$km: different choices of the mass and of the radius that are consistent with the fixed density value do not move the estimated strains, providing a numerical check of the goodness of the approximation made in Eq. \eqref{W caso generale}. 
This result indicates that, in the original FLE model, the strain developed by a spinning-down pulsar (for a given value of $\Omega\delta\Omega$) depends only on the average density of the star and on $L$: the independent choice of both $M$ and $R$ implies a degeneracy in the results. 

In a real NS the mass is a key parameter, and from the $M-R$ relation of realistic equations of state we know that more massive stars typically have smaller radii, implying a larger average density and smaller $W$, as can be seen in  Eq. \eqref{W caso generale}. 
In this sense, we can say that in the original FLE model heavier stars develop smaller strains during the spin-down, which is the expected behavior considering that the centrifugal force is less effective on more compact stellar configurations. However, the aim of the present work is a detailed (quantitative) parameter study and comparison between different models.

Therefore, as a final step, we study also how the changes of $L$ may affect $\alpha$. We found that the strain angle is a weak increasing function of the crustal thickness, namely a decreasing function of $L$ in the range $0.85\leq\, L\,\leq0.95$. 
For example, if we calculate $\alpha$ at $r=R'$ for a typical $M=1.4M_{\odot}$ NS as a function of $L$, we find that $\alpha(L=0.85)\simeq1.3\,\alpha(L=0.95)$.

\subsection{Cowling approximation}

It is possible to exploit the FLE model as a tool to estimate the importance of the so-called Cowling approximation, according to which the perturbation of the star's gravitational potential is neglected. Within the same original FLE scheme used in the previous section but assuming this further simplification, we can now rewrite the boundary conditions in Eq. \eqref{CONDIZIONI DI BORDO} as
\begin{align}
\begin{split}
& a-\frac{8}{21} A R^2 - \frac{B}{2 R^3} + \frac{8}{3} \frac{b}{R'^5}
=0
\\
& a-\frac{8}{21} A R'^{2} - \frac{B}{2 R'^3} + \frac{8}{3} \frac{b}{R'^5}
=0
\\
& -2f'(R) - \frac{ v_K^2 }{ c_t^2 } \frac{ f (R) }{R}+\frac{2}{3}\frac{ \Omega \delta \Omega }{c_t^2} R^2
= A R^2
\\
& -2  f'(R')
= AR'^2+\frac{B}{R'^3}  \, .
\label{CONDIZIONI DI BORDO COWLING}
\end{split}
\end{align}
Using the definition \eqref{SPOSTAMENTO FLE}, together with the solutions of the above equations, we obtain the corresponding displacement, that can be written in the dimensionless form introduced in Eq. \eqref{Spostamento Adimensionale}. The corresponding explicit form of the coefficients is given in Appendix B.
The simplest way to estimate the net effect of the perturbed gravitational potential is to neglect the terms containing the ratio $\chi$ and compare the displacement obtained with and without the Cowling approximation, indicated as $\boldsymbol{u}^{C}$ and $\boldsymbol{u}$ respectively. 
In the limit $\chi \, =\,0$ we find that
\begin{equation}
\frac{u_{r}}{u_{r}^{C}}
\, = \, 
\frac{u_\theta}{u_{\theta}^{C}}
\, = \, 
\frac{5}{2} \, + \, O\left( \, \chi^2 \,(1-L) \, \right) .
\label{eq:cowling}
\end{equation}
Therefore, in the FLE scheme, the displacements calculated with the Cowling approximation are $40\%$ of the ones calculated by considering also the gravitational potential perturbation (see also Appendix B).

\subsection{FLE model with $M$-$R$ relation from realistic equations of state}

Once that we have understood the main physical properties of the FLE model, we can study the strain developed in rotating NSs by using the mass-radius relation of two very different equations of state, the soft SLy \citep{douchin2001} and the stiff GM1 \citep{glendenning1991}. 

This  use of realistic EoSs, albeit still extremely approximate in this case of uniform density, links all the parameters of the star (i.e. $R$, $L$ and $\rho$) to its mass. This simplifies our parametric study and, at the same  time, gives an estimate of what we might expect in an astrophysical scenario where the mass is the key parameter that sets the stellar properties (once the EoS is known).

\begin{figure}
		\includegraphics[width=\columnwidth]{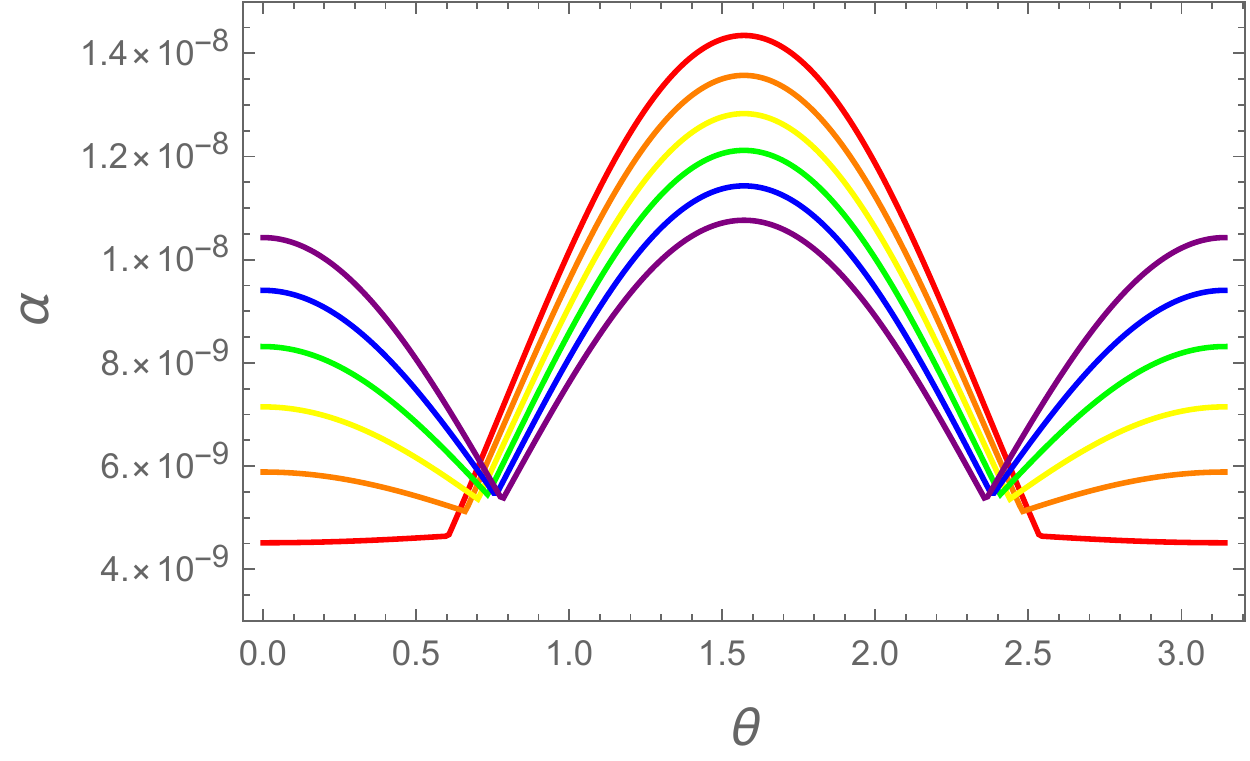}
    \caption{
    Strain angle $\alpha$ as a function of the colatitude for the original FLE model. 
    The strain is calculated for $M=1.4M_{\odot}$, with the SLy EoS, at different evenly spaced     values of $r$, from $r=R'$ (red) to $R$ (purple). Again, the maximum strain occurs at the core-crust interface on the equatorial plane.
    }
    \label{fig:STRAIN SLY VARI R}
\end{figure}

\begin{figure}
		\includegraphics[width=\columnwidth]{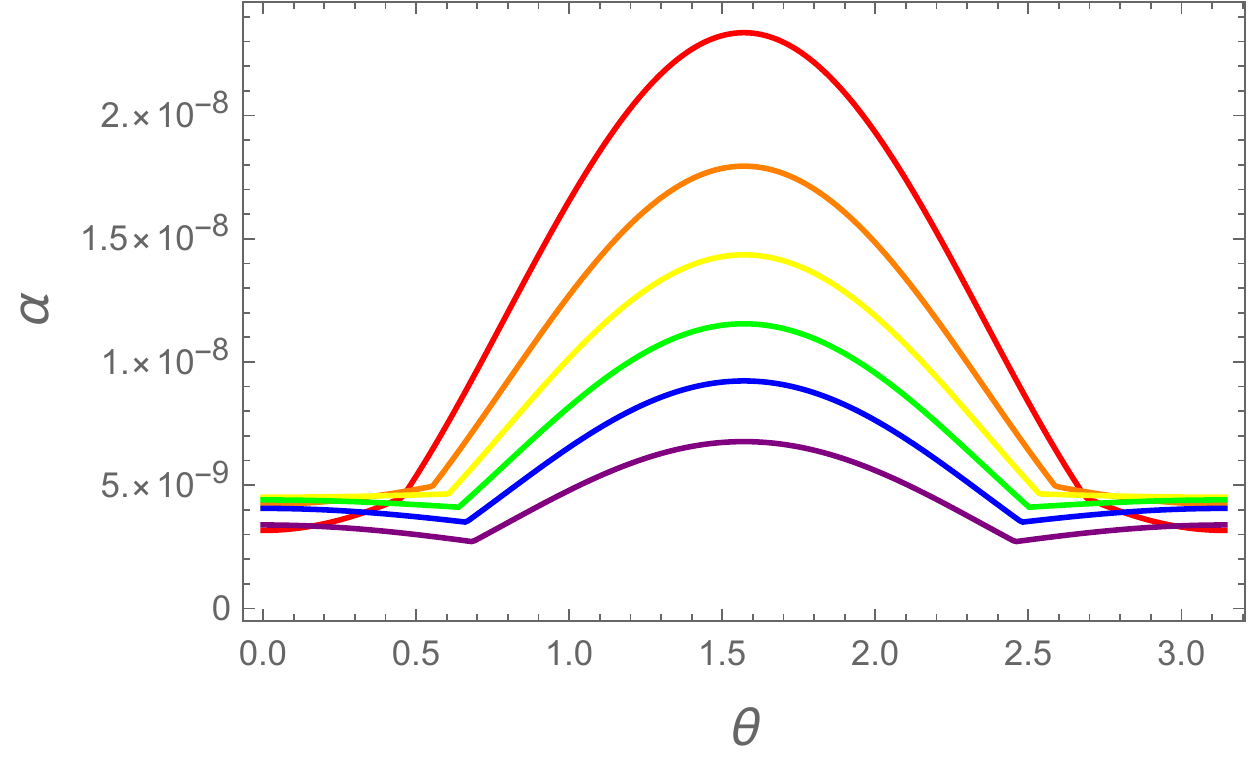}
		\caption{
     Strain angle $\alpha$ as a function of the colatitude for the original FLE model on a spherical shell of radius 
     $r=R^{'}$, i.e. where the strain angle reaches its maximum value. The structural parameters have been fixed by considering the SLy EoS, for different stellar masses: $M=1M_{\odot}$ (red), $M=1.2M_{\odot}$ (orange), $M=1.4M_{\odot}$ (yellow), $M=1.6M_{\odot}$ (green), $M=1.8M_{\odot}$ (blue), $M=2M_{\odot}$ (purple).
     }
    \label{fig:STRAIN SLY FLE VARIE MASSE}
\end{figure}

\begin{figure}
		\includegraphics[width=\columnwidth]{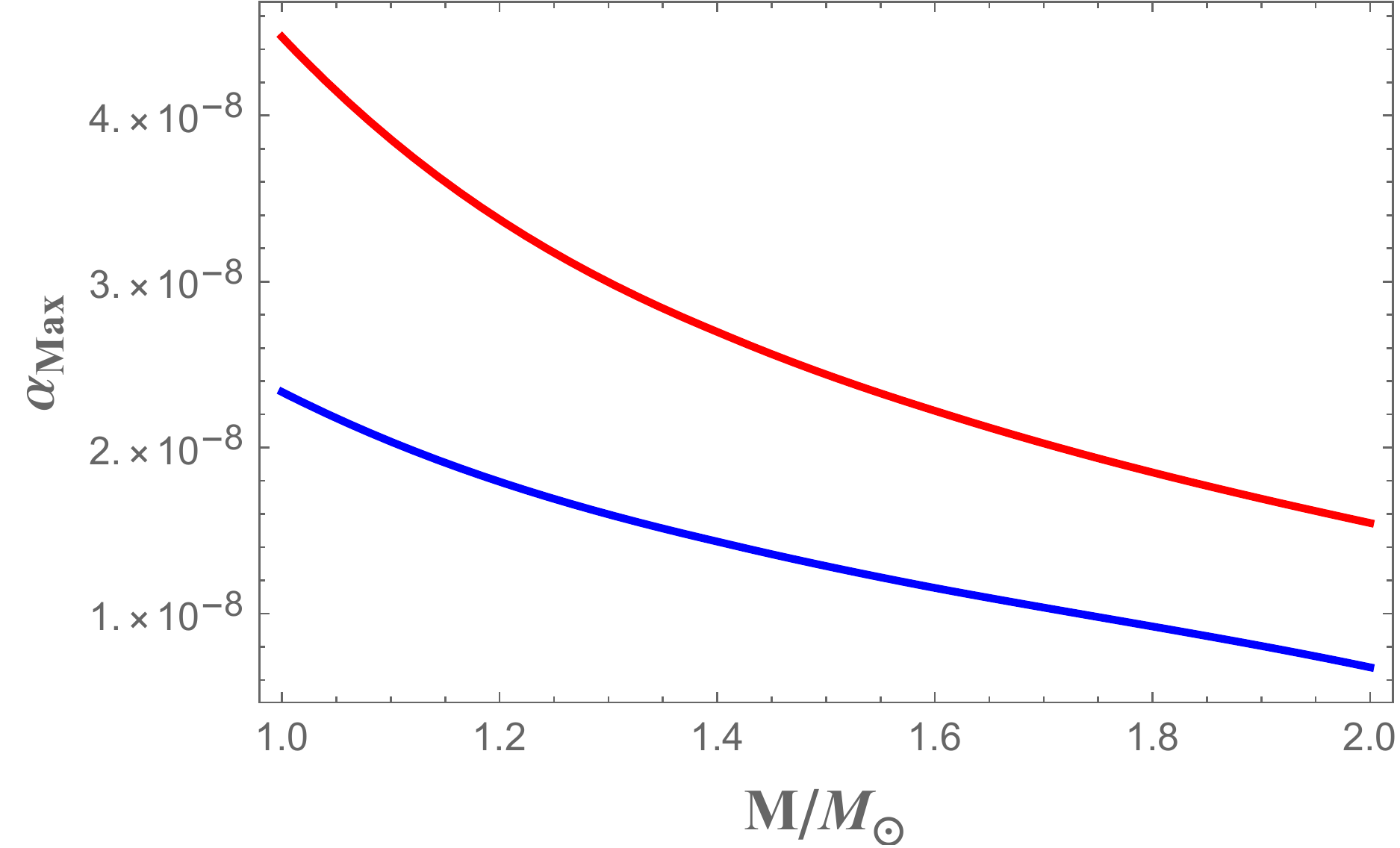}
		\caption{
    Comparison of the maximum values of the strain angle $\alpha_{max}$ (which always occurs at $r=R^{'}$ and $\theta=\pi/2$), obtained with the original FLE model, as function of the stellar mass. A comparison between the SLy EoS (blue) and GM1 EoS (red) is made for our benchmark value $\Omega\delta\Omega=1\,$rad$^2$/s$^2$. 
    The curves approach for higher masses as the crust thickness decreases and $R'$ gets closer to $R$; the GM1 line remains always well above SLy because a stiffer equation of state gives a thicker crust for the same mass.}
    \label{fig:STRAIN SLY vs GM1}
\end{figure}

As expected from the previous analysis, and shown in Figs \ref{fig:STRAIN SLY VARI R} and \ref{fig:STRAIN SLY FLE VARIE MASSE}, the strain is a decreasing function of the radius and of the mass. On the other hand, the comparison between different EoSs shows slightly different strains, as can be seen in Fig \ref{fig:STRAIN SLY vs GM1}: a stiffer equation of state gives larger maximum values of $\alpha$. 
Again, this is has to be expected from Eq. \eqref{W caso generale}; for the same stellar mass, a stiffer EoS gives a larger stellar radius, and thus a smaller compactness.

The most important information that can be extracted from Fig \ref{fig:STRAIN SLY vs GM1} is that the maximum strain value (of the order of $\alpha_{max} \sim 10^{-8}$) is three orders of magnitude smaller than the lowest allowed breaking strain ($\sigma_{max} \sim 10^{-5}$), and therefore it's unlikely that, according to the FLE model, the spin down between two subsequent glitches could deform the crust enough to break it: the only viable possibility is that the crust has to be always in a stressed state, near the failure threshold.
 
As already anticipated, all the values plotted in the figures have to be multiplied by $\Omega\delta\Omega$ in order to obtain the actual strain values of a specific pulsar. 
It is possible to give a rough estimate of the average spin-down $\delta\Omega$ that occurs in between two glitches of an active pulsar by considering the spin down rate $\dot \Omega<0$ and the average waiting time between glitches $\delta t$; we set $\delta\Omega\, =\, |\dot \Omega|\,\delta t$. 


In Table \ref{tab:table 1}, the specific values of $\Omega\delta\Omega$ are reported for a selection of pulsars with at least $10$ recorded events. Clearly, the most interesting pulsars for the present analysis are the ones with large values of the product $\Omega\, |\dot \Omega|\,\delta t $; the record holder is J0537-6910, followed by the Vela pulsar. 
As we can see, except for the Vela and J0537-6910, the rotational term can only decrease the values for the strain amplitude discussed above (that were all calculated for $\Omega\delta\Omega=1\,$rad$^2$/s$^2$). The strain remains well below the critical threshold even in the limit of very light stars with stiff equation of state. Furthermore, if we consider the highest current estimate of the breaking strain value $\sigma_{max} \sim 0.1$, we generally will not expect the crust to break via the spin down mechanism in the whole star life, as has been recently proposed by \cite{fattoyev2018}.

\begin{table}
	\centering
	\caption{
	The rotational parameter $\Omega\delta\Omega$ that sets the actual value of the average stress developed in between two glitches is given for a selection of pulsars with at least 10 glitches. Data are taken from the Jodrell Bank Glitch Catalogue \citep[\href{url}{www.jb.man.ac.uk/pulsar/glitches.html}, see also ][]{espinoza2011}.
	}
	\label{tab:table 1}
	\begin{tabular}{lccr} 
		\hline
		Pulsar Name	& $\Omega\delta\Omega$ [rad$^2$/s$^2$]\\
        \hline
J0537-6910	& 5$\pm$2 \\
J0631+1036	& 0.006$\pm$ 0.004\\
B0833-45 (Vela)	& 0.6$\pm$ 0.3\\
B1338-62	& 0.04$\pm$ 0.02\\
B1737-30	& 0.002$\pm$ 0.002\\
B1758-23	& 0.005$\pm$ 0.003\\
B1822-09	& 0.0002$\pm$ 0.0001\\
		\hline
	\end{tabular}
\end{table}

Finally, we can also compare the maximum strain angle using the original FLE approach ($\alpha_{FLE}$) with the one obtained by using the homogeneous model of Baym and Pines ($\alpha_{BP}$), where the star is described as an elastic, rotating, homogeneous spheroid. We choose $M=1M_{\odot}$, and calculate all the other quantities according to the SLy equation of state, since the use of a light and soft star should emphasize differences. 
Both $\alpha_{FLE}$ and $\alpha_{BP}$ are the evaluated for $r=R^{'}$ and $\theta=\pi/2$, where the strain angle is maximum. In this case we obtain very similar values for the two models:
\begin{align*}
& \alpha_{BP}=2.33\times10^{-8}
\\
& \alpha_{FLE}=2.55\times10^{-8}.
\end{align*}
This result leads us towards another further step: the study of a FLE-like model in which the crust and the core can have different average densities. 


\section{TWO-DENSITY MODEL}

The original FLE model provides a useful tool to estimate the deformation of a rotating NS in Newtonian gravity, but it is based on the strong assumption that the star must have the same constant density everywhere. 
In this section we show how to overcome this limitation, by using a self-consistent approach, where the neutron star is divided in two homogeneous layers representing the fluid core and the crust, with densities $\rho_{f}$ and $\rho_{c}$ respectively. 
As we will show, the self-consistency of the model becomes manifest in two additional conditions for the gravitational potential. Note, in fact, that contrary to the FLE model here one cannot use the knowledge of the gravitational potential of a perturbed homogeneous spheroid, but has to calculate it self-consistently by solving the perturbed Poisson equation. 

The present analysis is based on the more general result discussed by \citet{sabadini1996}, where it is shown that it is possible to develop and build analytical models containing a large number of layers as a description of auto-gravitating rocky planets. 
Here this set of ideas is adapted to the rotating neutron star problem. 
In particular, the present approach takes inspiration from the two-layer model firstly developed by \citet{sabadini82} in the context of viscoelastic planets (like Earth). 
The main equations and the global work scheme are summarized in Appendix A, while more technical details can be found in \citet{sabadini_book} and in the recent description of a class of more realistic (i.e. continuously stratified and auto-gravitating) neutron star models \citep{Giliberti2018}. 

In our two-density model, we find that the displacement $\boldsymbol{u}$ still has the same analytic form of the displacement given in Eq. \eqref{SPOSTAMENTO FLE}; this is not surprising as the main difference with respect to the original FLE model lies in the treatment of the boundary conditions. 
In fact, at $r=R'$ we have a finite density discontinuity between the core and the crust, a detail which has to be carefully incorporated into the analysis of the crust-core interface. 
As a consequence, if we write down the displacement $\boldsymbol{u}$ in the form of Eq. \eqref{Spostamento Adimensionale},  the four coefficients $\tilde{a}, \tilde{b}, \tilde{A}, \tilde{B}$ will be functions not only of the thickness $L$ and of $\chi$, but also of the density ratio 
\begin{equation}
d=\frac{\rho_{c}}{\rho_{f}}<1.
\label{definizione d}
\end{equation}
As for the previous model, a simplified form for the coefficients $\tilde{a}$, $\tilde{A}$,  $\tilde{b}$ and  $\tilde{B}$  is given in Appendix B
(the complete and exact form of the coefficients turns out to be much more complex with respect to the previous cases).

%

\subsection{A first comparison with the original FLE model}

We start by pointing out that the original FLE model can be obtained as a trivial limit $d = 1$ of our two-density model. 
In fact, imposing $d=1$ in our model, we calculate the resulting displacement $\boldsymbol{u}$ and the analogous one (i.e. by using the same values of $L$, $R$, $M$ and $\chi$) with the FLE model, $\boldsymbol{u}^{FLE}$. The ratio between the two gives
\begin{equation}
    \frac{u_r}{u^{FLE}_r}\, =\,\frac{u_\theta}{u^{FLE}_\theta}\, =\, 1 \,\,\mathrm{for}\,\,d=1 .
    \label{rapporto Sabadini FLE}
\end{equation}
In other words, our model can be seen as a complete generalization of FLE approach, accounting in a self-consistent way for two different density in the NS core and crust.

\begin{figure}
		\includegraphics[width=\columnwidth]{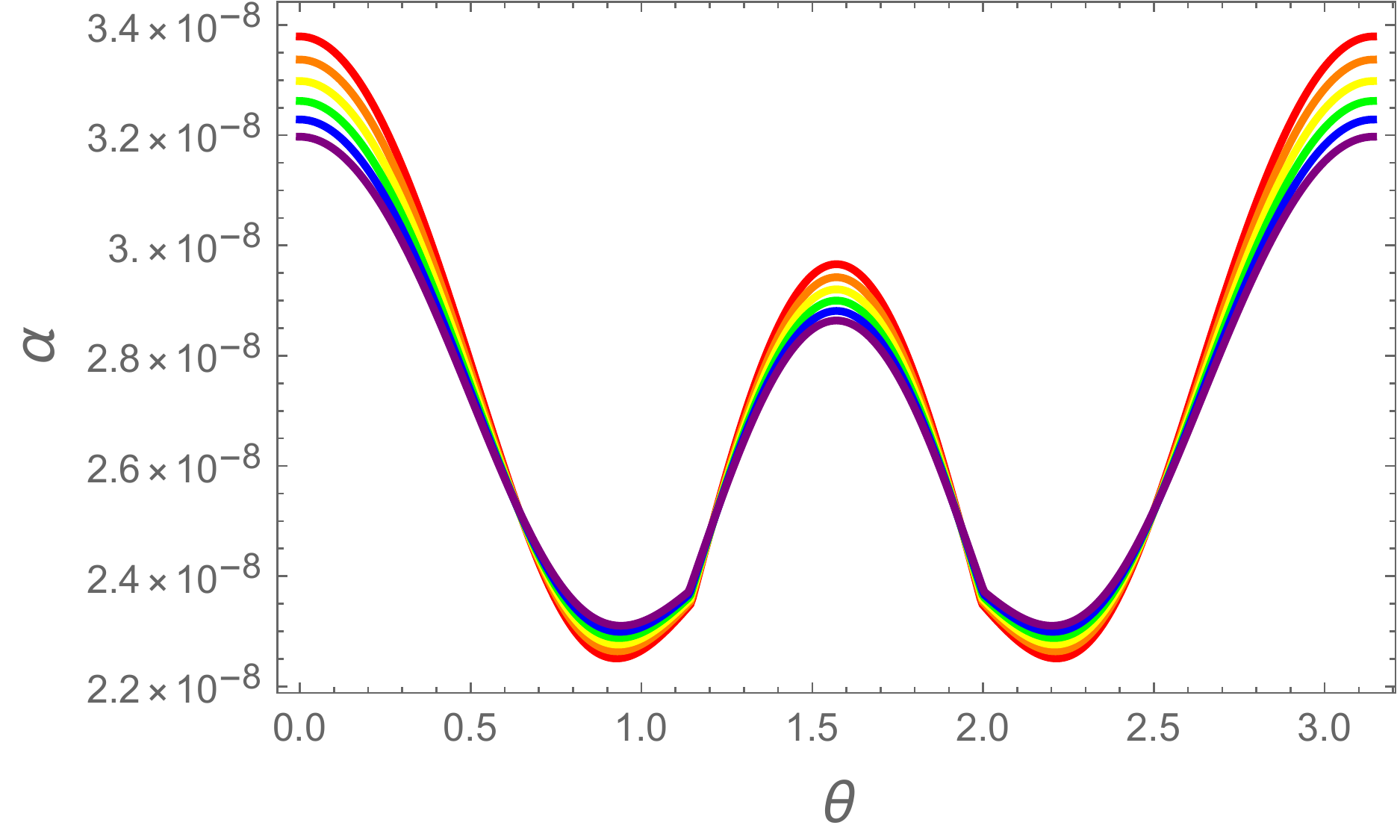}
    \caption{
    Strain angle as a function of the colatitude $\theta$ for the two-density model 
    and fixed benchmark values $M=1.4M_{\odot}$, $R=10\,$km, $L=0.95$, $\rho_f=6.6\times10^14$, $\rho_c=\rho_f/10$. 
    The strain angle is calculated for different values of the radius: $r=R$ (purple), $r=0.99 R$ (blue), $r=0.98 R$ (green), $r=0.97 R$ (yellow), $r=0.96 R$ (orange) and $ r=0.95R$ (red). 
    We used $\Omega\delta\Omega=1\,$rad$^2$/s$^2$.}
    \label{fig:STRAIN 2D TUTTO VARIABILE RAGGI DIVERSI}
\end{figure}

We now follow the same analysis done for the FLE model in the previous section, varying in turn one parameter while keeping the others fixed. Since the parameter space is rather large, we will vary several parameters at the same time by using a realistic EoS in the next subsection. 
However, as a preliminary example, we make a comparison with the FLE model by studying a situation similar to the one described in Fig \ref{fig:STRAIN TUTTO VARIABILE DIVERSI RAGGI}, which corresponds to a star of uniform density $\rho= 6.6\times10^{14}\,$g/cm$^3$ and mass $M=1.4M_{\odot}$.
In Fig \ref{fig:STRAIN 2D TUTTO VARIABILE RAGGI DIVERSI} we plot the strain for the two-density model for some fiducial values of the parameters involved, with $\rho_f\approx 6.6\times10^{14}\,$g/cm$^3$ and $d=0.1$, such that the total mass is still of $1.4\,M_\odot$.
Firstly, we note that the strain angle in this case is larger with respect to the FLE one. Furthermore, as the radial dependence of $\alpha(r,\theta)$ shows, the strain angle reaches its maximum value $\alpha_{max}$ at the crust-core interface. However, differently with respect to the FLE model, in this case the value of the strain is highest at the poles. 

As a final comparison, despite the fact that $\alpha$ is still a decreasing function of $L$, we note that the crust thickness has a even smaller impact on the strain value respect to the FLE model. In this case, in fact, we find, for a $1.4\,M_{\odot}$ NS, $d=1/10$,  $\alpha(L=0.85)=\alpha(L=0.95)\simeq1.03$.

\subsection{Realistic equations of state}

\begin{figure}
		\includegraphics[width=\columnwidth]{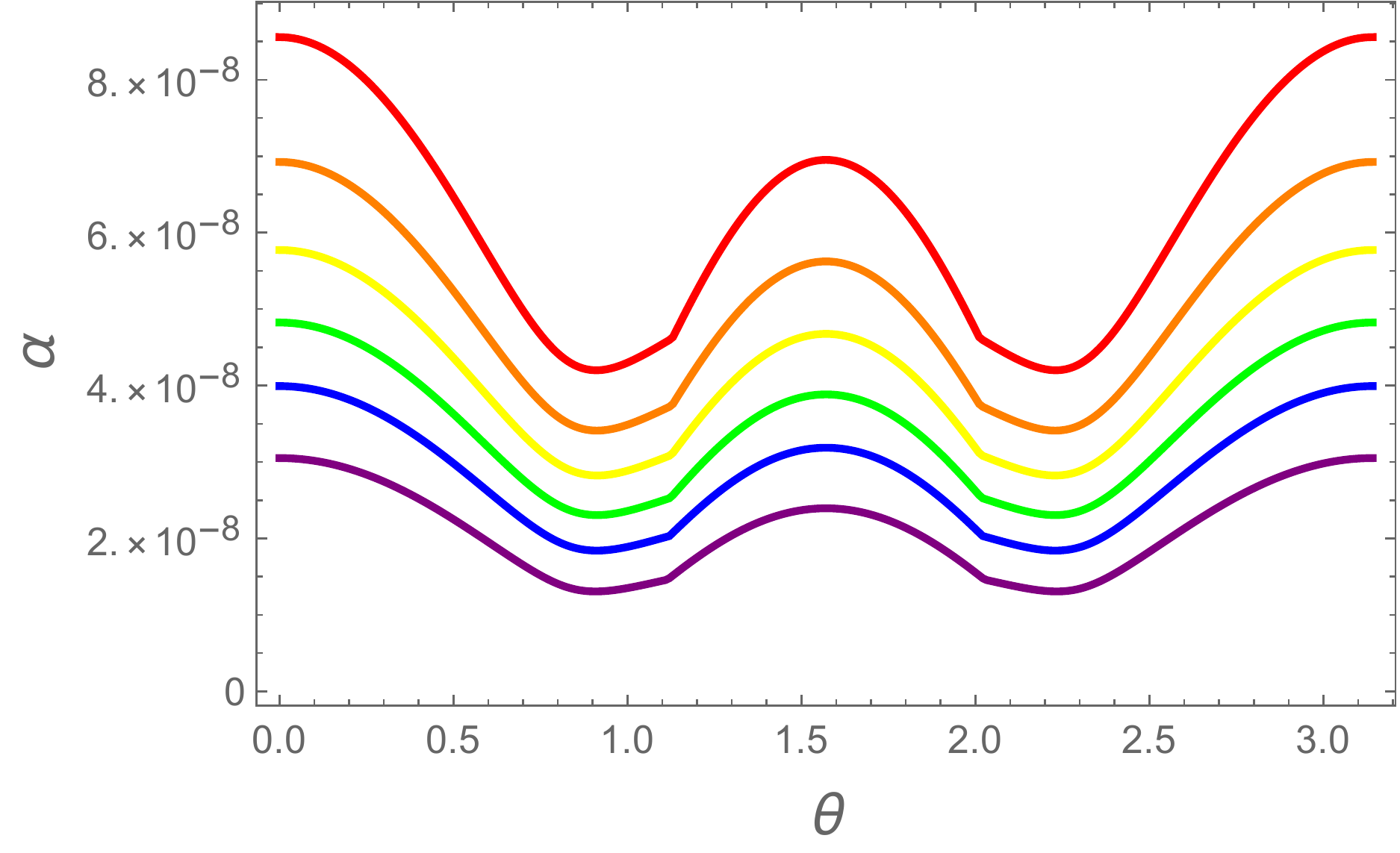}
    \caption{
    Strain angle at $r=R^{'}$ as a function of the colatitude $\theta$ for the two-density model and different masses: $M=1M_{\odot}$ (red), $M=1.2M_{\odot}$ (orange), $ M=1.4M_{\odot}$ (yellow), $M=1.6M_{\odot}$ (green), $M=1.8M_{\odot}$ (blue), $M=2M_{\odot}$ (purple). The SLy EoS has been used to fix all the structural parameters in terms of the mass. 
    }
    \label{fig:STRAIN 2D MASSE DIVERSE}
\end{figure}

As already done for the FLE model in section 3.3, we investigate the behaviour of the two-density model by imposing that not all the parameters present in the equations are free: they have to satisfy the constrain which arises by the fact that an EoS for the internal matter is related to a particular mass-radius relation. 
In order to give a stricter comparison with the FLE's model, and since the crust contains only a small percentage of the total stellar mass, we use this simple prescription 
\begin{equation*}
    \rho_{f}\simeq\frac{M}{4/3\pi R^3}.
\end{equation*}
On the other hand, the exact value of $d$, which definition is in Eq. \eqref{definizione d}, is given by the appropriate relation due to the particular EoS that has been chosen.
We start by considering the SLy equation of state; in Fig \ref{fig:STRAIN 2D MASSE DIVERSE} the strain angle at the core-crust interface is shown for different stellar masses. 
As expected, also in this case we have that the strain decreases when the total mass is increased: again, heavier stars have smaller radii and higher density, and are thus more difficult to deform. However, we highlight the new interesting feature that never arises by using the original FLE model: the maximum strain $\alpha_{max}$ is now at the poles. 
Forcing our the model to have $\rho_{f}=\rho_{c}$, we find, clearly, that the maximum strain is placed at the equator \eqref{rapporto Sabadini FLE}.
Therefore, the stratification (i.e. the presence of different layers with different densities), introduces a new degree of freedom into the model, so that it is possible to move the region of maximum stress away from the equator.

\begin{figure}
\includegraphics[width=\columnwidth]{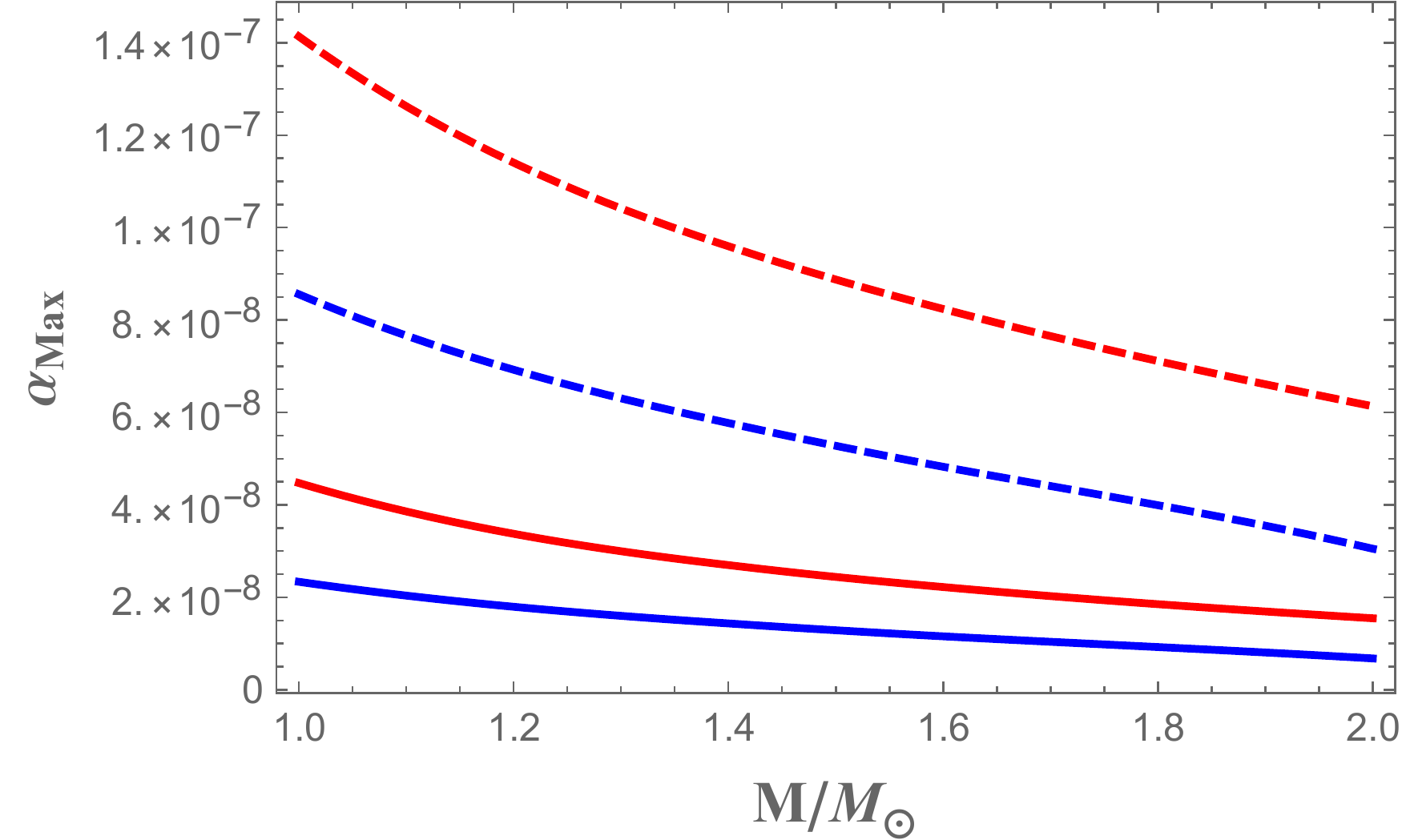}
\caption{
Maximum strain angle $\alpha_{max}$ (which occurs at the core-crust interface) as a function of the stellar mass for the FLE (solid curves) and for the two-density model (dashed curves). The red curves refer to the GM1 equation of state, blue curves to SLy.  
}\label{fig:STRAIN 2D vs FLE}
\end{figure}

Finally, despite the fact that with this model the strains are typically bigger than the FLE's ones, as can be seen in Fig \ref{fig:STRAIN 2D vs FLE}, we note that the maximum strain angle is still far even from the minimum breaking strain value of $\sim10^{-5}$. Therefore, the use of this more refined model confirms that, starting from an unstressed configuration, the deformation due only to the inter-glitch spin down is not large enough to trigger a starquake.

\section{CONCLUSION}


In this work we studied in details two different models describing the deformation of the crust of a rotating neutron star due to its spin-down: the original FLE model \citep{franco2000} and our two-density generalization, based on an adaptation of the scheme (valid for rocky planets) first proposed by \citet{sabadini82}. 
Despite the fact that both approaches are analytically solvable, the scheme proposed by \citet{sabadini82} has two main advantages: it is self-consistent for spherical auto-gravitating bodies (i.e. not only homogeneous ones) and partially accounts for stratification as it allows for two different densities in the core and in the crust.

Both schemes were introduced in the literature without a specific parameter study: the FLE model, for example,  was originally built for the study of pulsars precession, and solved only for a fiducial stellar configuration, while in the original work of \citet{sabadini82} the focus was on geophysical applications. Here, instead, we studied how the calculated strains vary by considering different stellar structures, where parameters such as the radius, the average density and the crust thickness are linked to the mass via an EoS. In order to parametrize our ignorance on the unknown equation of state for matter at supra-nuclear densities, an important part of the analysis has been performed employing  SLy and GM1 as a prototype of a soft and a stiff EoS respectively.

Three conclusions have been drawn. Firstly, all models (including the homogeneous limit of \citet{baym1971}) indicate that more compact stars are more difficult to deform (the strain scales with the inverse of the average density). Because of this quite general scaling, SLy is found to give smaller strains than GM1, as it gives rise to more compact configurations. 

Secondly, we found that the two-density model gives a strain angle that is about four times larger than the FLE one, although the dependence of the strain on various physical quantities is qualitatively the same in both models. This clearly indicates that the different density values of the core and crust is a fundamental aspect for the determination of the displacement and stress in rotating NS. 

As a third point, the maximum strain angle obtained using the two-density model (as shown in Fig \ref{fig:STRAIN 2D vs FLE}) differs for less than one order of magnitude according to the present analysis (the maximum strain for a neutron star of $M=1\,M_{\odot}$ is only about $\simeq2\div3$ times the one of a neutron star with $M=2\,M_{\odot}$, depending on the EoS used). 
Hence, it is not possible to conclude that the mass is a key parameter which clearly divides neutron stars in light objects that are easily  undergo crust failure and compact ones that are much more difficult to break.

Finally, using a group of pulsars that have been observed to glitch several times, we gave an estimate of the strain angle due to the spin-down between subsequent events. Both the FLE and the two-density model clearly indicate that starting from an initial unstressed state, it is not possible to develop enough strain to break the crust as frequently as glitches are observed. Therefore, the idea of a starquake as a trigger for the vortex avalanche in glitches of superfluid NSs is severely challenged, unless the crust is always in a state of great internal stresses (which is actually the case for the Earth and other rocky planets).
Since the two-density model gives typically larger and differently shaped strains with respect to the FLE one (e.g. the maximum strain can be at the poles), the study of a realistic stratification, without the assumption of incompressible medium, seems to be the obvious choice for future studies (see \citet{Giliberti2018}). 



\begin{acknowledgements}
The authors thank the PHAROS COST Action (CA16214). M. Antonelli acknowledges support from the Polish National Science Centre grant SONATA BIS 2015/18/E/ST9/00577, P.I.: B. Haskell.
\end{acknowledgements}


\begin{appendix}


\section{MODEL DERIVATION}

In this appendix we briefly outline the scheme used for the derivation of the two-density model \citep[see also ][]{sabadini_book}.
We assume that the neutron star has two homogeneous components, a fluid core and a solid crust, with density $\rho_{f}$ and $\rho_{c}$ respectively. 
Like in the original FLE model, the starting point is the differential equation which defines a static configuration 
\begin{equation}
\boldsymbol{\nabla\cdot T+F}=0,
\label{equazione equilibrio generale}
\end{equation} 
where $\boldsymbol{T}$ is the stress tensor and $\boldsymbol{F}$ is the body force acting on the star. Here we are considering only the effect of rotation, thus is natural to define the total (gravitational plus centrifugal) potential $\Phi$, which satisfies the Poisson equation
\begin{align}
\nabla^2 \Phi = - 4\,\pi\,G\,\rho+2\,\Omega^2 \, ,
\label{POISSON EQUATION}
\end{align}
where $\Omega$ is the angular velocity of the star.
By expanding around equilibrium, equations \eqref{equazione equilibrio generale} and \eqref{POISSON EQUATION} become
\begin{align}
\begin{split}
&
\boldsymbol{\nabla}\cdot\boldsymbol{\sigma} 
- 
\boldsymbol{\nabla}\cdot
\left(
\rho_0\,\boldsymbol{u}\cdot\boldsymbol{\nabla}\phi_0
\right)  
+
\boldsymbol{\nabla}\cdot(\rho_0\,\boldsymbol{u}) \, \boldsymbol{\nabla}\phi_0 
+...
\\
&\qquad\qquad\qquad\qquad\qquad\qquad\quad... -\rho_0\,\boldsymbol{\nabla}\Phi^{\Delta} 
= 0
\end{split}
\label{eq equilibrio espansa}
\\
&
\nabla^2 \Phi^{\Delta} = -4\,\pi\,G\,\boldsymbol{\nabla}\cdot(\rho_0\,\boldsymbol{u})+2\,\Omega^2 \, ,
\label{POISSON ESPANSA}
\end{align}
where the subscript $0$ indicates the unperturbed quantities, while $\boldsymbol{\sigma}$ and $\Phi^{\Delta}$ are the Cauchy tensor and the local variation of the potential\footnote{
In \cite{sabadini_book} a specific terminology is used. For a generic quantity $f$, the ``local increment'' $f^\Delta$ coincides with what is usually called Eulerian change \citep{shapiro_book}. 
On the other hand,  the Lagrangian changes of $f$ are dubbed ``material increments'' and are indicated by $f^\delta$. 
}  respectively. Using the assumption of constant shear and bulk modulus, an expansion in 
Legendre polynomials $P_{\ell}$ of all the physical quantities allows to rewrite the radial and tangential part  of Eq. \eqref{eq equilibrio espansa} and the Poisson equation \eqref{POISSON ESPANSA} as
\begin{equation}
\frac{\beta}{\rho_{0}}\partial_{r}\chi_{\ell}-\partial_{r}\left(gU_l\right)+g\chi_{\ell}-\partial_{r}\Phi_{\ell}+\frac{\mu}{\rho_{0}}\frac{\ell\left(\ell+1\right)}{r}H_{\ell}
=0
\label{radiale incomprimibile}
\end{equation}
\begin{equation}
\frac{\beta}{\rho_{0}}\chi_{\ell}-gU_{\ell }-\Phi_{\ell}+\frac{\mu}{\rho_{0}}\partial_{r}\left(rH_{\ell}\right) 
\,=\,0\, ,
\label{tangenziale incomprimibile}
\end{equation}
and
\begin{equation}
\nabla_{r}^{2} \, \Phi_{\ell}
=
-4\pi G\left(\rho_{0}\chi_{\ell}+U_{\ell}\partial_{r}\rho_{0}\right)
\, ,
\label{POISSON DEFINITIVA}
\end{equation}
where 
\begin{equation*}
\nabla_{r}^{2}=\partial_{r}+\frac{2}{r}\partial_{r}-\frac{\ell \left(\ell+1\right)}{r^{2}} \, .
\end{equation*}
The terms with the subscript $\ell$ are the expansion coefficient of the corresponding quantity in Legendre polynomials. Here, in particular, $U_{\ell}$ and $V_{\ell}$ are the radial and tangential displacement coefficient, i.e. $f(r)$ and $g(r)$ of equation \eqref{SPOSTAMENTO FLE} respectively.
In the above equations some quantities have been introduced: $\beta=\kappa+4/3\mu$ and $g$ is the gravitational acceleration at the initial state of hydrostatic equilibrium\footnote{The gravitational acceleration is defined as
\begin{equation*}
g=\frac{4\pi G}{r^{2}}\int_{0}^{r}r'^{2}\rho_{0}(r')dr'.
\end{equation*}}, while the scalars  $H_{\ell}$ and $\chi_{\ell}$ are defined as 
\begin{equation*}
H_{\ell}=\partial_{r}V_{\ell}+\frac{V_{\ell}-U_{\ell}}{r}
\, 
\end{equation*}
\begin{equation*}
\chi_{\ell}=\partial_{r}U_{\ell}+\frac{2}{r}U_{\ell}-\frac{\ell\left(\ell+1\right)}{r}V_{\ell}
\, .
\label{definizione chi}
\end{equation*}
In particular, the scalar functions $\chi_\ell$ are related to the relative volume change $\Delta$ via an expansion in Legendre polynomials as
\begin{equation}
\Delta 
\, = \, 
\boldsymbol{\nabla\cdot u}=\sum_{\ell=0}^{\infty}\chi_{\ell}P_{\ell}
\, .
\end{equation}
According to the incompressibility assumptions there are no volume changes $\Delta$ in our model. However, during the deformations, also incompressible materials must be able to react to isotropic stresses. We thus require that the bulk modulus $\kappa$ is infinitely large but in such a way that the material increment of the pressure, which can be expressed as $p^\delta=-\kappa \Delta$, remains finite \citep{sabadini_book}. Therefore, the coefficients $p_{\ell}$ of the expansion in spherical harmonics of $p^\delta$ remain finite when the formal limit
\begin{equation*}
p_{\ell}
\, = \,
- \lim_{ \substack{\chi_{\ell}\to 0 \\ \kappa \to \infty }} 
\kappa \, \chi_{\ell}
\,
\end{equation*}
is taken \citep{love59}. Using the incompressibility assumption $\chi_{\ell}=0$ into Eq. \eqref{definizione chi} gives a relation between the radial and the tangential displacements:
\begin{equation}
V_{\ell}=\frac{r\partial_{r}U_{\ell}+2U_{\ell}}{\ell \left(\ell +1\right)}
\, .
\label{relazione tra v e u}
\end{equation}
The quantity $H_{\ell}$ can thus be written as
\begin{equation}
H_{\ell}=\frac{\nabla_{r}^{2}\left(U_{\ell }r\right)}{\ell\left(\ell+1\right)}
\, .
\label{hl in funzione di u}
\end{equation}
%
%
%
Since in our model the layers are homogeneous (i.e. $\partial_{r}\rho_{0}=0$) the Eq. \eqref{POISSON DEFINITIVA} \emph{within} each layers becomes the Laplace equation ($\ell\geq1$)
\begin{equation}
\nabla_{r}^{2}\Phi_{\ell}=0 \, .
\label{laplace incomprimibile}
\end{equation}
As usual, we write the solution of Eq. \eqref{laplace incomprimibile} as 
\begin{equation}
\Phi_{\ell} = c_3 r^{\ell} + c_3^*r^{-\left(\ell+1\right)} \, 
\label{soluzione poisson}
\end{equation}
where $c_3$ and $c_3^*$ are constants of integration.
We underline that the absence of density perturbations within the layers does not implies that the gravitational perturbation $\Phi^{\Delta}$ is zero. 
Indeed, there are density jumps $\Delta\rho_{\lambda}$ between different layers at interfaces defined by $r=\lambda$ ,
\begin{equation}
\Delta\rho_{\lambda}=\rho_{0}\left(\lambda^{+}\right)-\rho_{0}\left(\lambda^{-}\right) \, .
\, 
\end{equation}
This means that we can write the density gradient as 
\begin{equation*}
    \partial_{r}\rho_{0}=(\rho_{c}-\rho_{f})\delta(r-R')
\end{equation*}
\begin{equation*}
    \partial_{r}\rho_{0}=-\rho_{c}\delta(r-R),
\end{equation*}
at the interfaces $r=R'$ and $r=R$, respectively. 

The introduction of the auxiliary quantity 
\begin{equation*}
\Gamma_{\ell}
=
-\frac{p_{\ell}}{\rho_{0}}-gU_{\ell}-\Phi_{\ell}
\, ,
\end{equation*}
allows us to reduce the Eqs \eqref{radiale incomprimibile} and \eqref{tangenziale incomprimibile} to
\begin{equation}
\partial_{r}\Gamma_{\ell}+\frac{\mu}{\rho_{0}}\frac{\ell\left(\ell+1\right)}{r}H_{\ell}
=0\, ,
\label{radiale riscritta}
\end{equation}
\begin{equation}
\Gamma_{\ell}+\frac{\mu}{\rho_{0}}\partial_{r}\left(rH_{\ell}\right)
=0 \, .
\end{equation}
These two equations can be combined into 
\begin{equation}
\nabla_{r}^{2} \, \Gamma_{\ell}=0,
\end{equation}
which has the solution
\begin{equation}
\Gamma_{\ell}=-\frac{\mu}{\rho_{0}} \, c_{1} \,r^{\ell}
-
\frac{\mu}{\rho_{0}} \, c_{1}^{*} \, r^{-\ell-1} 
\, ,
\end{equation}
where the quantity $\mu/\rho_{0}$ have been inserted for convenience and $c_1$, $c_1^*$ are two constants. 
Inserting this solution in \eqref{radiale riscritta}  and using the relation \eqref{hl in funzione di u} we obtain a differential equation for $U_{\ell}$: 
\begin{equation}
\nabla_{r}^{2} ( U_{\ell} \, r )
=
c_{1} \, \ell \, r^{\ell}
-
c_{1}^{*} \,(\ell+1) \, r^{-\ell-1}. 
\end{equation}
Solving this equation and using \eqref{relazione tra v e u} we finally get the radial and tangential displacements as 
\begin{equation}
U_{\ell}
=
c_{1}\frac{\ell\:r^{\ell+1}}{2\left(2\ell+3\right)}+c_{2}r^{\ell-1}+
c_{1}^*\frac{\left(\ell+1\right)r^{-\ell}}{2\left(2\ell-1\right)}+c_{2}^*r^{-\left(\ell+2\right)},
\label{u incomprimibile generale}
\end{equation}
\begin{multline}
V_{\ell}
=
c_{1}\frac{\left(\ell+3\right)r^{\ell+1}}{2\left(2\ell+3\right)\left(\ell+1\right)}+c_{2}\frac{r^{\ell-1}}{\ell}+...\\
...+c_{1}^*\frac{\left(2-\ell\right)r^{-\ell}}{2l\left(2\ell-1\right)}-c_{2}^*\frac{r^{-\left(\ell+2\right)}}{\ell+1}
\, .
\label{v incomprimibile generale}
\end{multline}
We remind that the centrifugal potential can be expanded in Legendre polynomials as
\begin{equation}
\phi^{C}\left(r,\theta,\varphi\right)
=
\phi_{0}^{C}\left(r\right)P_{0}\left(\theta\right)+\phi_{2}^{C}\left(r\right)P_{2}\left(\theta\right)
\, ,
\label{ESPANSIONE POTENZIALE CENTRIFUGO}
\end{equation}
where
\begin{equation}
\phi_{0}^{C}\left(r\right)=-\frac{\Omega^{2}r^{2}}{3}
\, ,
\end{equation}
and
\begin{equation}
\phi_{2}^{C}\left(r\right)
=
\frac{\Omega^{2}r^{2}}{3}
\,.
\end{equation}
All the harmonic coefficients with $\ell \neq 0, 2 $ are zero. Moreover, the coefficient with $\ell=0$ is suppressed by the request of incompressibility. Therefore, in the case of deformations induced by the centrifugal force we have only the harmonic contribution corresponding to $\ell=2$. In this scenario is easy to see that the above expression for the radial and tangential displacement \eqref{u incomprimibile generale} and \eqref{v incomprimibile generale} have the same form of the ones given in Eq. \eqref{SPOSTAMENTO FLE}.
Note that, differently from the FLE model, where there are only four constants ($a$, $A$, $b$ and $B$), now six coefficients need to be determined ($c_1$, $c_2$, $c_1^*$, $c_2^*$, $c_3$, $c_3^*$): two new coefficients come from the self-consistent treatment of the Poisson equation and have been introduced in Eq \eqref{soluzione poisson}. 
In order to fix these constants we have to impose the opportune boundary conditions at the interface between layers.

\subsection{Boundary conditions}
The boundary condition can be easily written by using the expansion in spherical harmonics of the material incremental stress, that is
\begin{equation}
\boldsymbol{\sigma}\cdot\boldsymbol{e}_{r}=\sum_{\ell}\left(R_{\ell}P_{\ell}\boldsymbol{e}_{r}+S_{\ell}\partial_{\theta}P_{\ell}\boldsymbol{e}_{\theta}\right),
\end{equation}
where $\boldsymbol{e}_{r},\boldsymbol{e}_{\theta}$ 
are the usual spherical unit vector. We call $R_{\ell}, S_{\ell}$ radial and tangential stress respectively.
We require the continuity of the radial stress and that the tangential stress must be zero both at the star's surface $r=R$ and at the core-crust boundary $r=R'$
\begin{align}
 R_{\ell}(R^{+})\,& =\, 0 \, 
\label{R esterno}
\\
 R_{\ell}(R^{+})\,& =R_{\ell}(R'^{-})
\label{R interno}
\\
 S_{\ell}(R^{+}) \,& = \, 0 \,
\label{S esterno}
\\
 S_{\ell}(R'^{+}) \,& = \, 0 \, .
\label{S interno}
\end{align}
In fact, the fluid core and the vacuum outside the star cannot support shear stress; moreover, pressure is zero for $r\geq R$.
%
%
We can add other two conditions for the potential, by introducing the \emph{potential stress} 
\begin{equation}
Q_{\ell}
=
\partial_{r}\Phi_{\ell}+\frac{\ell+1}{r}\Phi_{\ell}+4\pi G\rho_{0}U_{\ell}
\, .
\end{equation}
Starting from the Poisson equation \eqref{POISSON EQUATION}
and using the Gauss theorem in a pillow box placed at a radius $r=R$ and $r=R'$ respectively, we can write the boundary conditions in a compact form:
\begin{align}
 Q_{\ell}(R^{+})\,& =\,Q_{\ell}(R^{-})  
\label{Q esterno}
\\
 Q_{\ell}(R'^{+})\,& =\,Q_{\ell}(R'^{-}).
\label{Q interno}
\end{align}
In particular, 
%
%
using the spherical harmonic expansion \eqref{ESPANSIONE POTENZIALE CENTRIFUGO} of the centrifugal potential
\begin{equation}
\Phi_{\ell}^{C}\left(r\right)=\Phi_{\ell}^{C}\left(R\right)\left(\frac{r}{R}\right)^{\ell},\,\,\ell>0,
\end{equation}
equation \eqref{Q esterno} can be explicitly written as
\begin{equation*}
\partial_{r}\Phi_{\ell}\left(R^{-}\right)+\frac{\ell+1}{r}\Phi_{\ell}\left(R^{-}\right)+4\pi G\rho_{0}\left(R^{-}\right)U_{\ell}\left(R^{-}\right)=
\end{equation*}
\begin{equation}
=\frac{2\ell+1}{R^+}\Phi_{\ell}^{C}\left(R^{+}\right).
\label{q esterno}    
\end{equation}
%
%
%
%
%
%
The six boundary conditions (\ref{R esterno}, \ref{S esterno}, \ref{Q esterno}, \ref{R interno}, \ref{S interno}, \ref{Q interno}) fix the coefficients, giving us the analytical displacements, stresses and potential.

\section{Coefficients}

The displacement in Eq. \eqref{SPOSTAMENTO FLE} can be rearranged in the form \eqref{Spostamento Adimensionale}, that highlight in the pre-factor the main physical quantities of the problem. As defined in the main text, $R$ is the stellar radius, $R'$ the core-crust interface radius,  $c_t=\sqrt{\mu/\rho_c}$ is the transverse speed of shear waves and $v_K=\sqrt{GM/R}$ is the Keplerian velocity. 
We also recall the definition of two useful dimensionless parameters that have been used in the text: $\chi=c_t/v_K$ and $d=\rho_c/\rho_f<1$. 
Since the parameter $L=R'/R$ spans from about $0.86$ to $0.95$ when ``realistic'' EoS are used, we can write $L=1-q$ and expand the coefficients $\tilde{a}$, $\tilde{b}$, $\tilde{A}$, $\tilde{B}$ and $Q$ up to the second order in $q$.

\subsection{FLE model}

The displacements of the FLE model are given in Eq. \eqref{Spostamento Adimensionale}, where the coefficients have to be fixed by considering the boundary conditions. Their explicit form is

\begin{align}
	\tilde{a}& = 280 \left( 13 q^2 - 7 q + 2 \right)\nonumber
	\\
    \tilde{b}& = -5 \left( 643 q^2 - 232 q + 37 \right)\nonumber
    \\
    \tilde{A}& = 280 \left( 15 q^2 - 13 q + 5 \right)
    \\
    \tilde{B}& = - 560 \left( 70 q^2 - 27 q + 5 \right) / 3\nonumber
    \\
    Q / v_K^2 & = -35 \,  \left( q^2 \left( 240 \chi^2 - 109 \right) + q \left( 48 - 60 \chi^2 \right) - 11 \right) \, .\nonumber
    \label{Q FLE}
\end{align}

\subsection{Cowling approximation}

If we use the Cowling approximation, discussed in Sec 3.2, we have different parameters with respect to the FLE case:

\begin{align}
    \tilde{a}^C & =560 \left(13 q^2-7 q+2\right)\nonumber
    \\
    \tilde{b}^C & =-10 \left(643 q^2-232 q+37\right)\nonumber
    \\
    \tilde{A}^C & =560 \left(15 q^2-13 q+5\right)
    \\
    \tilde{B}^C & = - 1120 \left( 70 q^2 - 27 q + 5 \right) / 3\nonumber
    \\
    Q^C / v_K^2& =-175 \, \left( q^2 \left( 96 \chi^2 - 109 \right) + q \left( 48 - 24 \chi^2 \right) - 11 \right)\nonumber
    \, .
\end{align}
It is now east to check that Eq. \eqref{eq:cowling} is valid when the (very small) terms proportional to $q^2 \chi^2$ and $q \chi^2$ are neglected.

\subsection{Two-density model}

As discussed in Sec 4, the displacements for the two-density model have the same analytic form given in Eq. \eqref{Spostamento Adimensionale}. In this case, the coefficients which appear into the explicit solution of the displacements are given by
\begin{align*}
\tilde{a}\,=\,&
q^2 \big[ 840 d^2 \left(53 \chi^2-8\right) -8400 d \left(4 \chi^2-1\right) \nonumber
\\
&-1680 \big]+ \nonumber
\\
& +q \left[-12600 d^2 \chi^2+840 d^2+6720 d \chi^2-840 d\right]+\nonumber
\\
& +1680 d^2 \chi^2 \nonumber
\\ 
\tilde{b} \, = \,&
q^2 \big[ -15 d^2 \left(1603 \chi^2-78\right)+225 d \left(64\chi^2-7\right)+ \nonumber
\\
& +405 \big]+\nonumber
\\
& +q \left[-15 d^2 \left(6-360 \chi^2\right)-1920 d \chi^2+90 d\right] +\nonumber
\\
& -555 d^2 \chi^2\nonumber
\\
\tilde{A}\,=\,&
q^2 \big[15 d^2 (111 -728 \chi^2 )+15 d  (608 \chi^2-147 ) + \nonumber
\\
& +540 \big] +\nonumber
\\
& +q \left[ -15 d^2  \left(15-256 \chi^2\right)-2280 d \chi^2+225 d\right] +\nonumber
\\
& -600 d^2 \chi^2
\\
\tilde{B}\,=\,&
280 \, q^2 \left[ 3 d^2 \left(58 \chi^2-1\right)+d \left(6-104 \chi^2\right)-3 \right]+\nonumber   
\\
&  +280 \, q (16-43 d) d \chi^2 +\nonumber
\\
& +1400 d^2 \chi^2\nonumber
\\
Q/v_K^2\,=\,&
\frac{1}{2}q^2 \big[
-1890 d^3 ( 26 \chi^2-3)+210 d^2 (167 \chi^2-36)+
\\ 
&-630 \, d \, (14 \chi^2-3) \big]+\nonumber
\\
& +\frac{1}{2}q  \left[210 d^3  (73 \chi^2-3)-5250 d^2  \chi^2+630 d^2  \right]+\nonumber
\\
& -1155 d^3 \chi^2 \, . \nonumber
\end{align*}
The exact form of these coefficients is much more complex, here expressions have been truncated to the second order in $q$, which is the relative crust thickness.

\end{appendix}



\end{document}